\documentclass[12pt]{iopart}

\usepackage{graphicx}
\begin{document}

\title[Two-step phase-shifting interferometry]{Two-step phase-shifting interferometry for phase-resolved imaging from a spatial light modulator}

\author{Lark E. Bradsby, Andrew A. Voitiv and Mark E. Siemens}

\address{Department of Physics and Astronomy, University of Denver, Denver, CO, USA}
\ead{msiemens@du.edu}
\vspace{10pt}
\begin{indented}
\item[]June 2024
\end{indented}

\begin{abstract}
We demonstrate two-step phase-shifting interferometry (holography) of complex laser modes generated by a spatial light modulator (SLM), in which the amplitude and phase of the signal are determined directly from measurements of phase-shifted interferograms. 
The reference and signal beams are generated and phase-controlled with a single composite hologram on the SLM and propagated collinearly. This requires no additional optics and leads to measurements that are more accurate and less prone to noise, which we demonstrate with collinearly-referenced measurements of various Laguerre-Gaussian modes and structured images. 
\end{abstract}

\vspace{2pc}

\section{Introduction}

Optical holography is a well-established technique for arbitrary control over the amplitude and phase structure of a propagating laser mode. This powerful approach, akin to arbitrary wavefunction control in the quantum fluids analogy \cite{carusotto-2013}, enables many applications such as generating optical vortices \cite{carpentier-2008}, tailoring optical tweezers \cite{padgett-2011}, and 3D holographic imaging \cite{shaked-2009}. Characterizing the amplitude of holographically-generated images 
is straightforward with a camera, and phase measurements are usually achieved with interferometry.
In a single interference measurement between a signal wave $S$ and reference wave $R$ at reference phase shift $\delta$, the measured interference $I_{R+S,\delta}$ is
\begin{equation}
I_{R+S,\delta} = I_R + I_S + 2 \sqrt{I_R I_S} \cos(\phi-\delta),
\label{eq:interference}
\end{equation}
where $I_R$ and $I_S$ are the intensities of the reference-only and signal-only waves, and $\phi$ is the phase of the signal relative to the (unshifted) reference. The cosine cannot be uniquely inverted, so phase cannot be directly recovered from a single collinear interferogram.

Phase-shifting interferometry (PSI) \cite{creath-1985}, also called phase-shifting holography because of its applications in imaging \cite{smith-1969, picart-2015}, is a standard technique for recovering the phase of an optical signal wave without the phase sign ambiguity that is intrinsic to a single interference measurement \cite{lewis-1993, flores-2020, creath-1988}. In PSI, multiple interferograms are recorded at different phase steps $\delta$, and then those measurements are used to calculate the amplitude and phase of the signal. 

There are many variants of PSI that use various numbers of phase-shifted interferograms: five-step \cite{schwider-1983}, four-step \cite{yamaguchi-1997, meng-2006, andersen-2019}, three-step \cite{lewis-1993}, and even just two-step \cite{santoyo-1988, liu-2009,tahara-2021}. Two-step PSI requires fewer images than the others, which is advantageous for fast phase measurements.The two-step PSI work of \cite{liu-2009} allowed for an arbitrary (known) reference phase shift between the interferograms; follow-on work \cite{meng-2006} used a $\pi/2$ phase step for direct quadrature measurements (see section S.1 of the Supplementary Information for a demonstration that $\delta = \pi/2$ is optimal). Quadrature PSI generates phase-resolved images by measuring the real and imaginary parts of the mode at each pixel \cite{strobel-1996}, while
direct phase PSI directly measures the phase and amplitude \cite{flores-2020}. Interferometry and the phase delays are usually achieved using an external interferometer, which can lead to additional noise from vibrations and air currents \cite{andersen-2019}. 

\begin{figure}
    \centering
    \includegraphics[width=\linewidth]{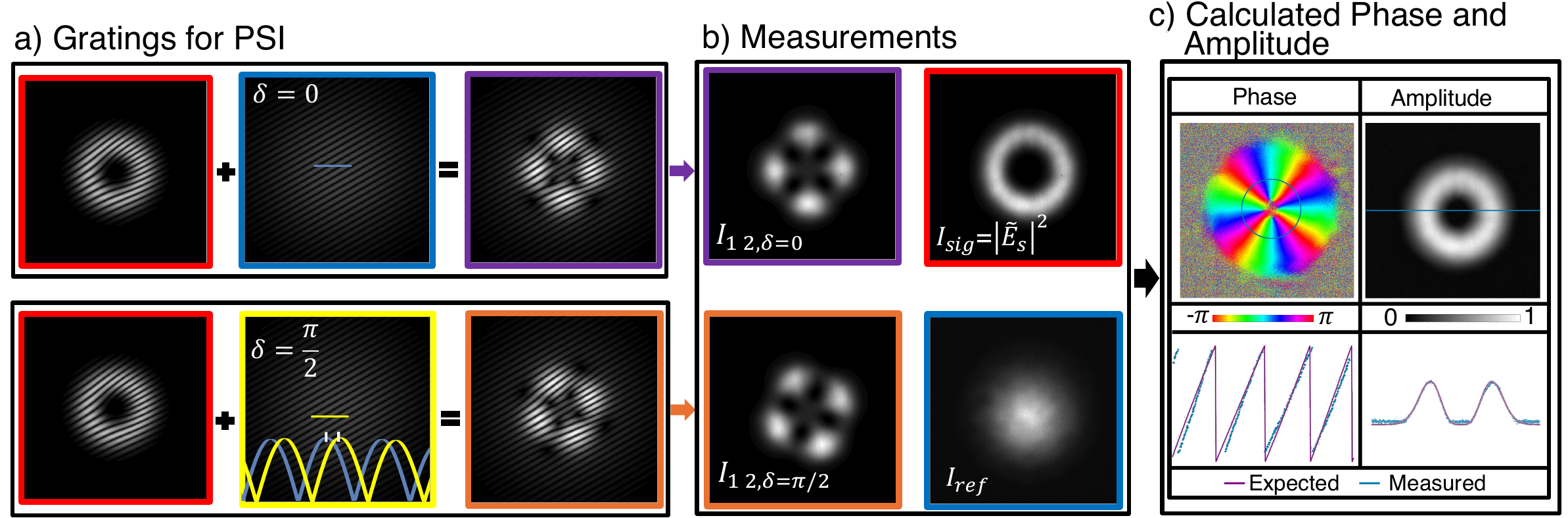}
    \caption{Diffraction gratings for an $\ell=4$ Laguerre-Gaussian (LG) mode with associated measurements and combined calculations.  a) Programmed composite diffraction holograms used for generation of signal and reference beams and combined for generation of  interference beams. Yellow and blue lineouts are from the short horizontal lines, and illustrate the shift in the hologram lines caused by the $\delta=0$ (blue) or $\delta=\pi/2$ (yellow) phase on the reference mode. b) Raw PSI measurements of an $\ell=4$ LG mode that where used to recover phase and amplitude information. c) Calculated phase and amplitude displayed on a 2D chart as well as line-outs with comparison to expected values.} 
    \label{fig:Concept}
\end{figure}

In this paper, we demonstrate collinear two-step phase-shifting interferometry.   Our approach uses 
composite holograms as shown in Figure \ref{fig:Concept}a on a single spatial light modulator (SLM) to generate both the desired image and a phase-controlled reference beam; these modes then propagate collinearly and interfere on a camera where measurements are taken (Figure \ref{fig:Concept}b). We can then directly calculate phase images shown in Figure \ref{fig:Concept}c from just two interference measurements. We use this approach to compare the quadrature vs direct-phase PSI algorithms, and confirm that the direct-phase approach exhibits lower noise and higher accuracy. We demonstrate this method using  complex image structures as well as Laugarre-Gaussian (LG) modes \cite{allen-1992}. 

\section{Theory of quadrature vs direct-phase PSI}

 We begin by showing how the phase can be directly measured in two-step PSI. Consider the interference described by Equation \ref{eq:interference} for the case of an unshifted reference ($\delta=0$) and solving for the relevant trigonometric function for the phase $\phi$:
\begin{eqnarray}
\cos\phi = \frac{(I_{R+S,\delta=0} - I_R - I_S)}{(2 \sqrt{I_R I_S})}. \label{eq:cos}
\end{eqnarray}
Repeat for the interference with a quadrature-shifted reference:
\begin{eqnarray}
\cos(\phi-\pi/2) = \sin \phi = \frac{(I_{R+S,\delta = \pi/2} - I_R - I_S)}{(2 \sqrt{I_R I_S})}. \label{eq:sin}
\end{eqnarray}
Equations \ref{eq:cos} and \ref{eq:sin} identify $\cos\phi$ and $\sin\phi$, which are used in quadrature PSI to calculate the real and imaginary parts of the complex signal field $\tilde E_S$: Re$[\tilde E_S] = \sqrt{I_S} \cos \phi$ and Im$[\tilde E_S] = \sqrt{I_S} \sin \phi$ (via Equations \ref{eq:cos} and \ref{eq:sin}). 
The direct-phase approach differs in that these results are combined to directly calculate the phase by combining them in the form of $\tan\phi = \sin{\phi}/\cos \phi$. Solving for $\phi$, we find
\begin{eqnarray} 
\phi  =  \textrm{atan} \left[ I_{R+S,\delta = \pi/2} - I_R - I_S \hspace{.05in} , \hspace{.05in}  I_{R+S,0} - I_R - I_S \right], \label{eq:tan}
\end{eqnarray}
where atan$\left[x,y\right]$ is the two-component version of the arctangent function that accounts for quadrant information. 

In practice with real data, one more step is needed. Unlike other similar methods of PSI such as four-step PSI, two-step PSI does not automatically subtract off the background (dark counts picked up by the CCD sensor of a camera) for each measurement, so the background should be manually subtracted in two-step PSI. Subtracting from each image an additional background measurement of the intensity with all beams blocked, $I_{BG}$, the phase can be calculated as:
\begin{equation}\label{eq:final}
\phi = \textrm{atan} \left[ I_{R+S,\delta=\pi/2} - I_R - I_S + I_{BG}, \hspace{.025in}  I_{R+S,\delta = 0} - I_R - I_S  + I_{BG}\right]
\end{equation}

The complex signal $\tilde E_S$ is obtained by combining the phase measurement $\phi$ of Equation \ref{eq:final} with the amplitude obtained from the signal intensity measurement (with the background subtracted): 
\begin{equation}
    \tilde E_S = \sqrt{I_S-I_{BG}} \exp{[i \phi]}.\label{eq:finalE}
\end{equation}
Compared to the quadrature PSI approach, the direct-phase approach of Equation \ref{eq:final} has two advantages that become evident with experimental data that contain noise: 1.) The amplitude of the signal is obtained straight from the intensity image, and 2.) it avoids the ``divde by zero'' problem of Equations \ref{eq:cos} and \ref{eq:sin} when the reference or signal are weak, which leads to exaggerations of noise wherever the reference or signal are near zero (such as outside the beam, or the center of a dark spot in the signal). This issue is not as apparent without background subtraction, but foregoing background subtraction leads to misleading phase measurements of the background phase \cite{zhang-2023, servin-2009}.

\section{Experimental demonstration} 
\begin{figure} [h!]
    \centering
    \includegraphics[width=.6\linewidth]{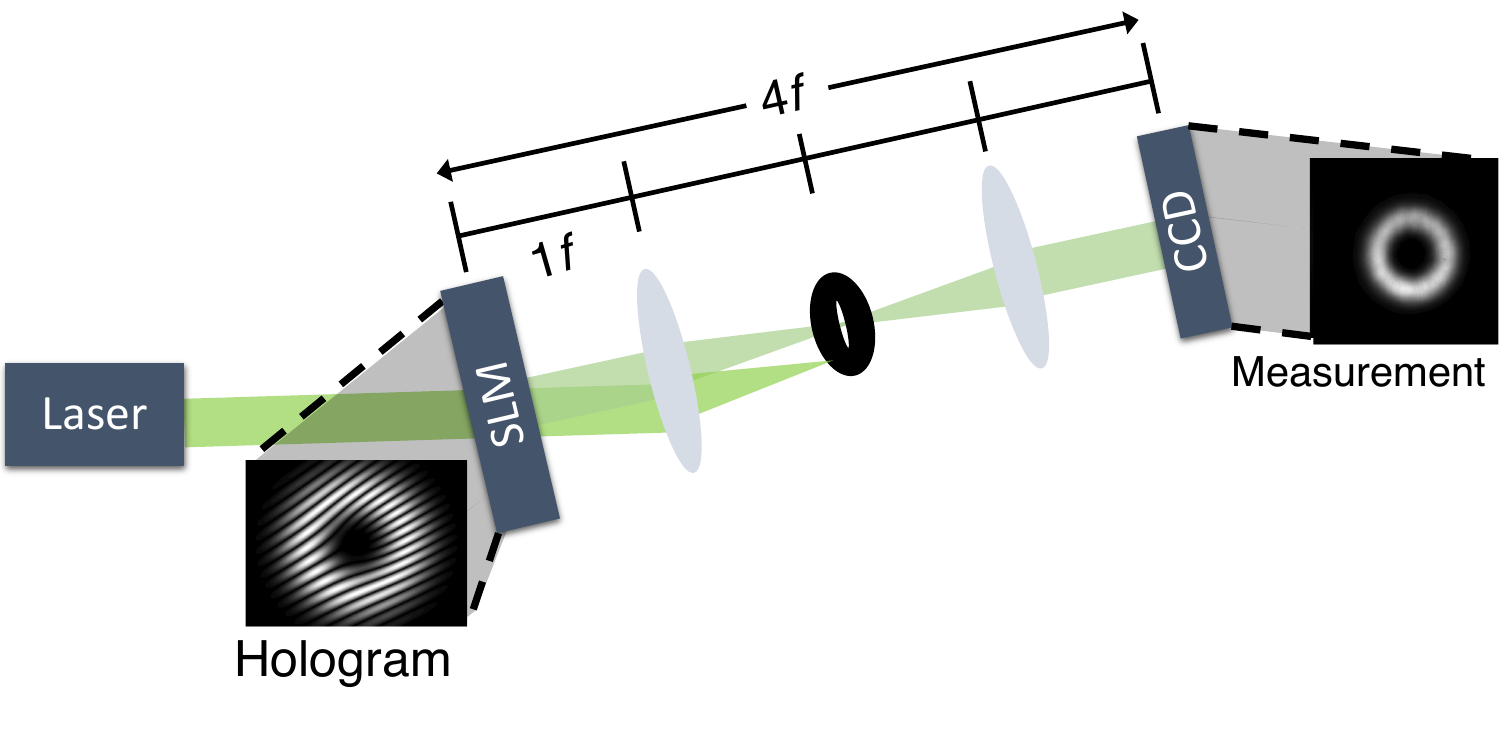}
    \caption{ Experimental schematic. A  532 nm laser 
    is incident on a computer controlled spacial light modulator (SLM). The first-diffracted order is passed through a 4f-imaging system with two lenses and and iris that blocks all other diffracted orders.  A CCD is placed in the imaging plane.
    }
    \label{fig:schematic}
\end{figure}

For experimental demonstration, a collimated laser with wavelength of $532$ nm is passed through a computer-controlled liquid crystal panel from a modified projector, used as a transmission SLM \cite{huang-2012}, as shown in Figure \ref{fig:schematic}. Accurate amplitude calibration of this SLM is essential for the 2-step PSI algorithm to work correctly; our calibration scheme is described in section S.3 of the Supplementary Information. 
The first-diffracted order is passed through a 4f-imaging system consisting of two lenses with an iris in the Fourier plane between them to block modes other than the first-order diffraction from the SLM. A camera is placed at the image plane where we will measure amplitude and phase images.

For demonstration, Laguerre Gaussian (LG) and other complex signal modes were generated. LG modes are characterized by a doughnut-shaped intensity profile with a central dark core and a helical phase structure that gives rise to orbital angular momentum. To implement the signal and  reference beams with a controlled phase shift between them, we make composite diffractive holograms to generate both beams, with independent control of each. This approach allows us to directly control the relative phase $\delta$ between the signal and reference modes, and propagate the modes in a collinear, common-path geometry that is robust to drift and noise from vibrations and air currents. 

The reference hologram is generated by interfering a plane wave with constant phase $\delta$ ($\exp[i \, \delta]$) with a tilted plane wave 
($\exp[i \, (\sqrt{3} \, x + y ) / 2]
$) to make the diffraction grating. Then a Gaussian  mask ($\sqrt{2/\pi} \, \exp[- (x^2+y^2)/w^2]$) is multiplied onto the grating to control the amplitude, yielding a hologram (transmission function) in the form:
\begin{equation}
    T_{R}(\delta) = \Bigg| \exp\left[i \, \delta\right]+\exp\left[i \, ( \frac{\sqrt{3}}{2} x + \frac{1}{2} y)\right] \Bigg| \times \sqrt{\frac{2}{\pi}}\exp\left[- \frac{x^2+y^2}{w^2}\right]
\end{equation}
The signal hologram is generated by interfering the signal mode, $\exp[i \, \phi$], with the same tilted plane wave and then multiplying the diffraction grating by the amplitude of the desired signal mode:
\begin{equation}
    T_{S}(\tilde E_S)=\Bigg|\exp\left[i \, \phi\right]+\exp\left[i \, ( \frac{\sqrt{3}}{2} x + \frac{1}{2} y )\right] \Bigg| \times \textrm{abs}[\tilde E_S]
\end{equation}

\noindent where from Equation \ref{eq:finalE}: $\phi =$ arg[$\tilde E_S$]. Signal and hologram gratings are used separately for measurements of $I_S$ and $I_R$, and are also combined in the form $T_{S}(\tilde E_S)+ T_{R}(\delta=0)$ and $T_{S}(\tilde E_S)+ T_{R}(\delta=\pi/2)$ for measurements of $I_{R+S, \delta = 0}$ and $I_{R+S, \delta = \pi/2}$. The relative amplitude of each grating is meaningful and affects the phase measurement, so the gratings are collectively normalized before measurements by calculating the maximum value among all gratings and dividing each grating by this maximum value.  We show example gratings, as well as specific equations for implementing the $\ell$=4 Laguerre-Gaussian LG mode gratings, in Figure S.3 of the Supplementary Information.

\section{Results}
\begin{figure}[h!]
\centering
\includegraphics[width=.58\linewidth]{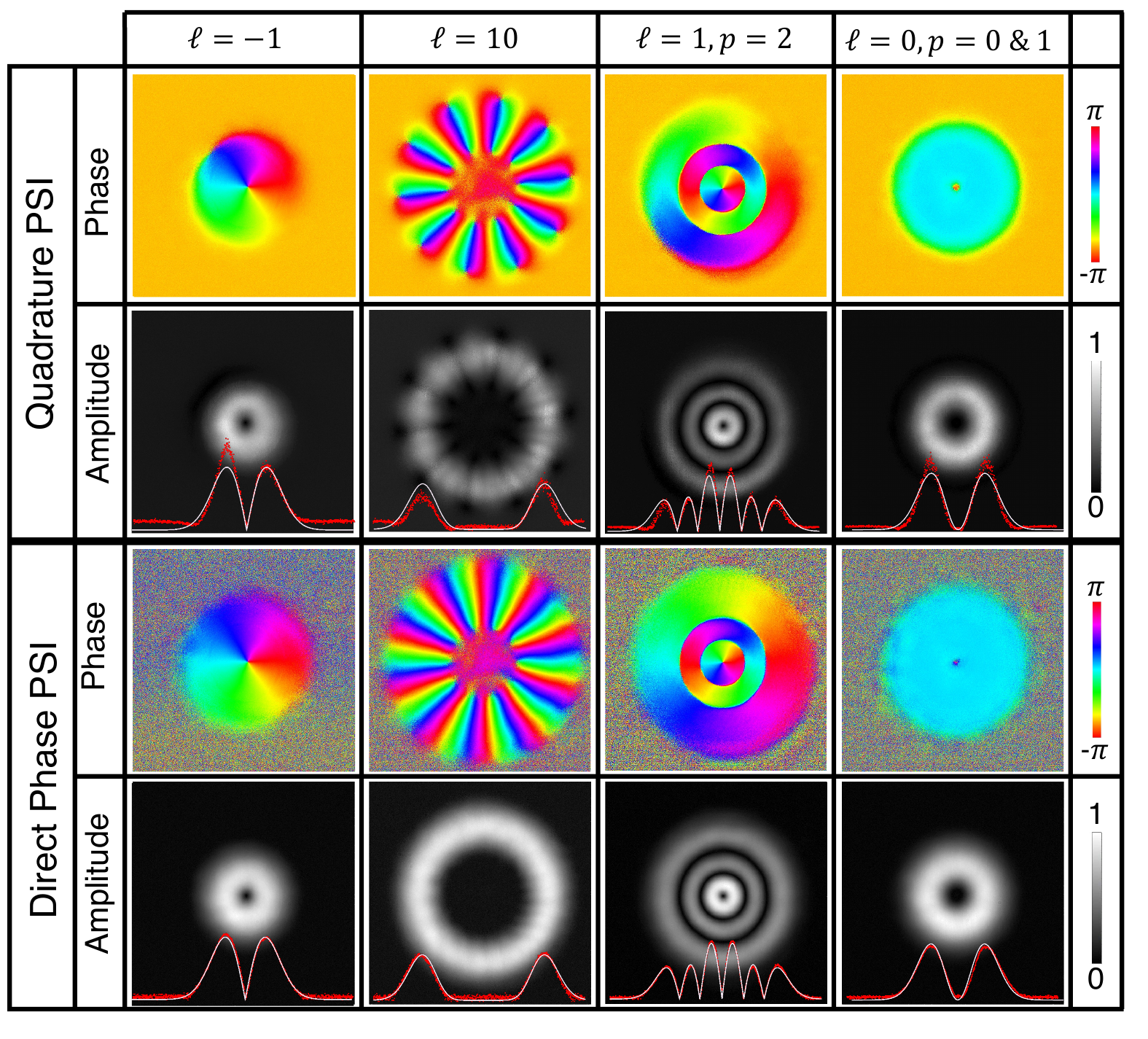}
\caption{Experimental amplitude and phase measurements of several complex Laguerre-Gaussian modes using Equations \ref{eq:cos} and \ref{eq:sin} for quadrature phase shifting interferometry and Equation \ref{eq:final} for direct-phase PSI. Slices though the center of the amplitude images are shown as red lines, and the white lines indicate the programmed expectation. The direct phase measurement has much better fidelity. The yellow background on the quadrature phase measurements is not physical and comes from unsubtracted background. 
}
 \label{fig:LGmeasure}
\end{figure}
Figure \ref{fig:LGmeasure} shows  Laguerre-Gaussian (LG) modes measured using both  direct-phase and quadrature PSI. Direct-phase PSI was calculated using Equations \ref{eq:final} and \ref{eq:finalE}
quadrature PSI was compiled using Equations \ref{eq:cos} and \ref{eq:sin} as well as
 Re$[\tilde E_S] = \sqrt{I_S} \cos \phi$ and Im$[\tilde E_S] = \sqrt{I_S} \sin \phi$. The comparison shows that the direct PSI produces images with better fidelity in both amplitude and phase. The amplitude  images via the quadrature method are characterized by inauthentic peak shapes, particularly shown by the lineouts showing slices through the center of each measured mode. In the case where the background is subtracted (not shown), amplitude images depict significant noise in the center and corners where the amplitude should be zero. In the phase measurements, the direct-phase PSI has a multicolored noise background (similar to the white noise of television static) that is meaningful because it represents regions where the amplitude is so small that phase information is not definite, while in contrast the phase measurement taken with the quadrature method has a solid (orange) phase that nonphysically assigns a definite phase value.

\begin{figure}[h!]
\centering
\includegraphics[width=.58\linewidth]{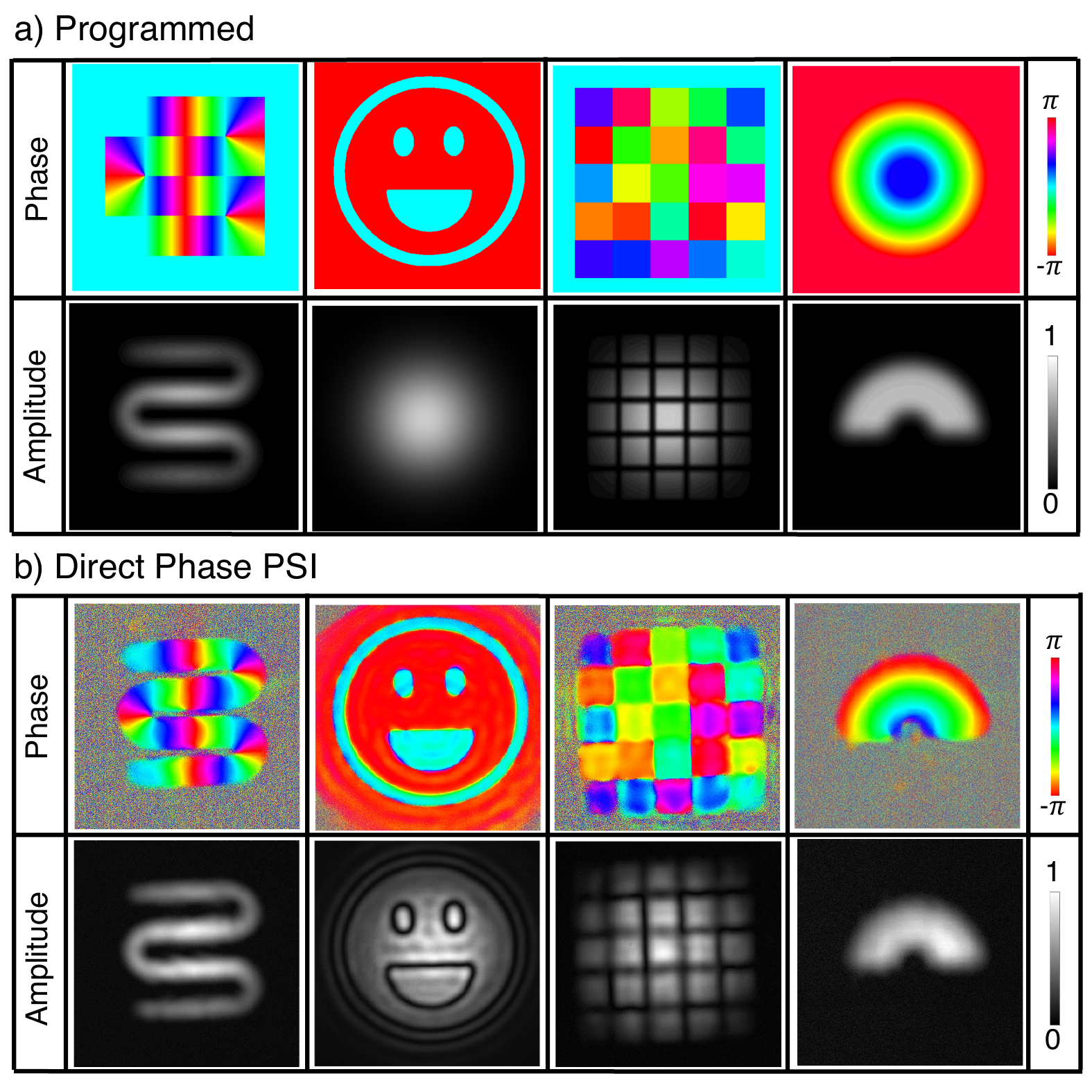}
 \caption{Phase and amplitude of complex image structures that were a) programmed onto the SLM via holography, and b) measured in the image plane via direct-phase measurement.} 
 \label{fig:images}
\end{figure}

To test and demonstrate phase-resolved imaging with direct-phase PSI, we generated four complex modes, shown in Figure \ref{fig:images}a, and programmed these as signal holograms to display on an SLM. Additionally, two phase-stepped combined reference and signal interferometric gratings were generated and resulting beams were measured as well. Measurements were acquired using the setup described above. Results of direct-phase PSI applied to measurements in the image plane of the SLM are shown in Figure \ref{fig:images}b. The first column shows an amplitude squiggle with a phase gradient, which exhibits excellent phase reproduction outside of the dark regions. The second column shows a phase-only smiley phase; the phase of these measurements is very accurate, while the amplitude exhibits artifacts arising from the sharp edges in the programmed-mode amplitude and the finite numerical aperture of the lenses in the imaging system. The third column shows a grid of random phases, and again the measured phase is quite accurate. The last column shows a phase rainbow, which is also well-reproduced in the experimental measurement. Note that in all of these measurements, regions with zero or negligible amplitude exhibit random phase in the measurement, which is consistent with the meaning of an undefined phase for zero amplitude.

\begin{figure}[hbt!]
    \centering
    \includegraphics[width=.58\linewidth]{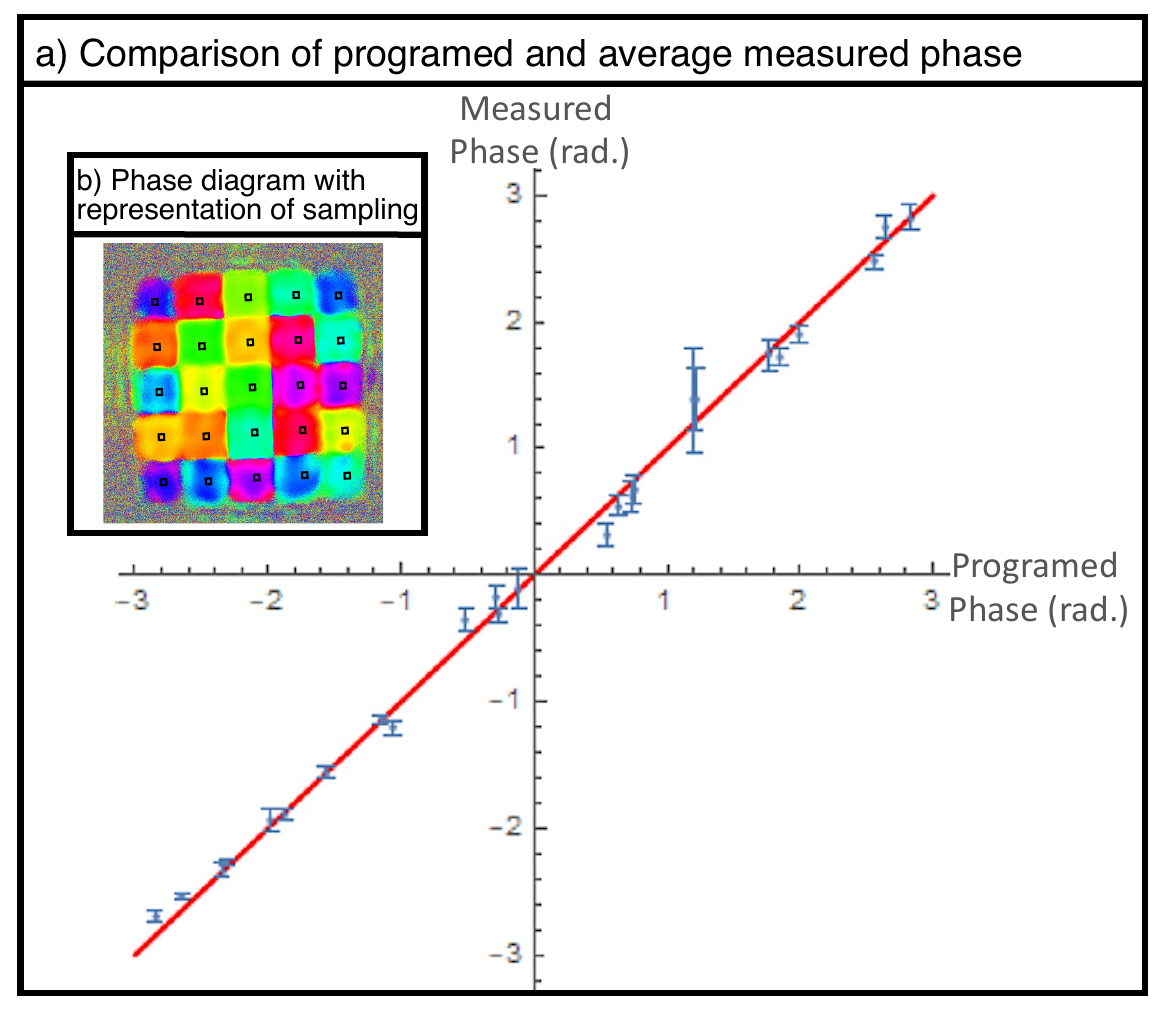}
    \caption{Graphical representation of experimental phase accuracy in the patchwork phase measurement shown in Figure 3 in the paper. a) Comparison graph of measured and programmed phase values of the patchwork phase diagram, b) patchwork phase diagram with the location of the 10 by 10 pixel square used for averaging.}
    \label{fig:PatchworkPhase}
\end{figure}

To test the accuracy of the phases measured in the laser mode of patchwork phases, we performed a quantitative comparison between the measured phase and the programmed phase in each square. Using the directly-measured phase image, notably without any phase roll, the average of each patch was determined by averaging a $10 \times 10$ pixel square at the center of each patch, as shown in Figure \ref{fig:PatchworkPhase}b. Figure \ref{fig:PatchworkPhase}a shows the quantitative comparison: each data point corresponds to a single set of sampled phase measurements: the x-value is the programmed phase, and the y-value is the average measured phase in the $10\times10$ pixel square at the center. Error bars show the standard deviation of the phase measured in those 100 pixels. The red line in Fig. \ref{fig:PatchworkPhase}a depicts the ideal of a 1:1 match between measured and programmed phases; the measured data matches this prediction within experimental errors with a linear regression coefficient is $0.99$.

\section{Conclusion} 

In this paper, we demonstrated a method of collinear two-step phase-shifting interferometry using composite diffractive holograms on a single spatial light modulator (SLM). By generating both the desired laser mode and a phase-controlled reference to propagate collinearly, we were able to  calculate phase images from just two interference measurements. Our comparison of the direct quadrature and direct phase PSI algorithms indicate that the direct-phase approach exhibits lower noise and higher accuracy. Because two-step PSI requires fewer image acquisitions than other methods, and is generated on one optical path, this technique can be applied in dynamic phase-resolved imaging.

\section*{Acknowledgments}
Funding:  NSF DBI 1337573 and DBI 1919541.

\section*{Data availability statement} 
Data underlying all results presented are available from the authors upon reasonable request.

\section*{Disclosures.} The authors declare no conflicts of interest.

\section*{References}

\bibliography{2PSI}

\pagebreak
\setcounter{equation}{0}
\setcounter{figure}{0}
\setcounter{table}{0}
\setcounter{page}{1}
\setcounter{section}{0}

\title[Supplementary Document: Two-step phase-shifting interferometry]{Supplementary Document: Two-step phase-shifting interferometry for phase-resolved imaging from a spatial light modulator}

\author{Lark E. Bradsby, Andrew A. Voitiv and Mark E. Siemens}

\address{Department of Physics and Astronomy, University of Denver, Denver, CO, USA}
\ead{msiemens@du.edu}
\vspace{10pt}
\begin{indented}
\item[]June 2024
\end{indented}

\section{Arbitrary-shift vs quadrature vs direct-phase 2-step PSI}

Two-step phase-shifting interferometry (PSI) was originally proposed for a reference phase shift $\delta$ between the two steps, where %that 
$\delta$ could be an arbitrary phase step
[9]. Subsequent work specialized this to quadrature PSI, for which $\delta = \pi/2$ [6]. It is intuitive that $\delta=\pi/2$ is optimal since minimal sensitivity is obtained in the limit of $\delta \rightarrow n \pi$, where $n$ is an integer.

\begin{figure}
    \centering
    \includegraphics[width=.8\linewidth]{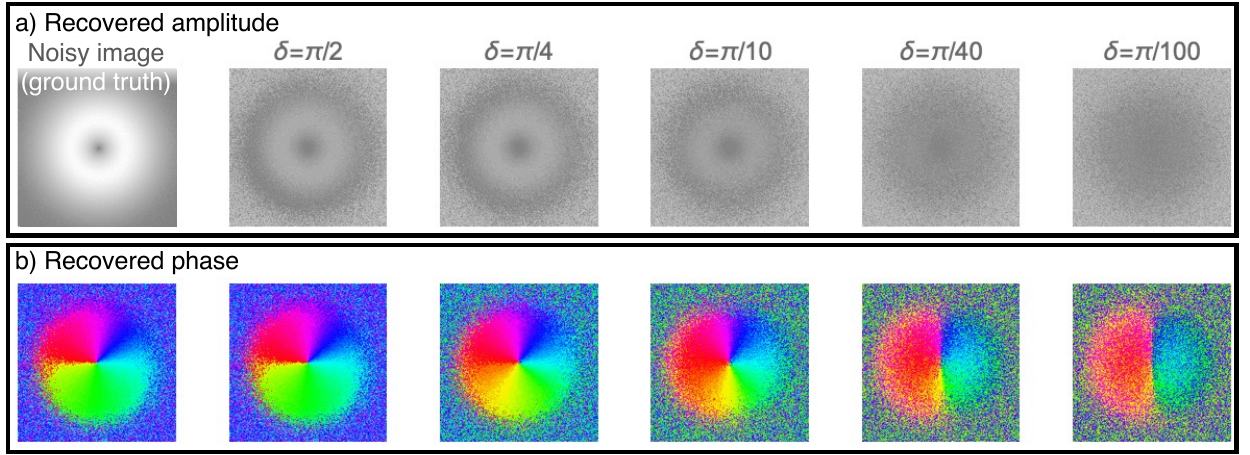}
    \caption{Calculated a) amplitude and b) phase with 2-step phase-shifting interferometry (PSI) with different phase shifts $\delta$. Ground truth is at left. }
    \label{fig:ArbitraryVsQuadraturePSI}
\end{figure}

For analytical models with ideal modes, there is no advantage to a particular value of $\delta$. The advantage of $\delta=\pi/2$ is seen only in calculations with real-world (i.e., noisy) data. This is seen in Figure \ref{fig:ArbitraryVsQuadraturePSI}, which shows computationally-modeled two-step PSI results for different $\delta$ values. In the calculation, the signal is a 2D table built from the complex field values for an ideal $\mathrm{LG}_{\ell=+1}^{p=0}$ mode.  Reference waves are also numerated onto a 2D table: one with $\delta = 0$ and another with $\delta$ as specified in Figure \ref{fig:ArbitraryVsQuadraturePSI}. Signal and reference data are individually normalized so that their transverse-integrated intensities are equal. Positive Gaussian noise equal to 2\% of the peak value is added to both signal and reference tables, and interferograms are formed. Then the process of Ref. [9] is applied to calculate the amplitude and phase from the noisy interferograms.

It is clear from Figure \ref{fig:ArbitraryVsQuadraturePSI} that $\delta=\pi/2$ gives the best reproduction of both amplitude and phase. However, it is also notable that, while the phase for $\delta = \pi/2$ is fully accurate in how it renders the noisy \emph{phase}, the amplitude is incorrectly calculated by two-step PSI. All of the calculated images appear significantly dimmer than the raw noisy amplitude because the amplitude calculation by the two-step PSI process is very susceptible to \emph{divide by zero} errors where the signal or reference are near zero. This leads to noise spikes in the corners of the image; in this data, the noise was up to $75\times$ larger than the maximum true signal level. The reduced brightness in the PSI-recovered amplitude images in  Figure \ref{fig:ArbitraryVsQuadraturePSI} is from auto-normalization to these noise spikes. The \emph{direct-phase PSI} proposed in this paper obviates this problem.

\begin{figure}[hbt!]
    \centering
    \includegraphics[width=.7\linewidth]{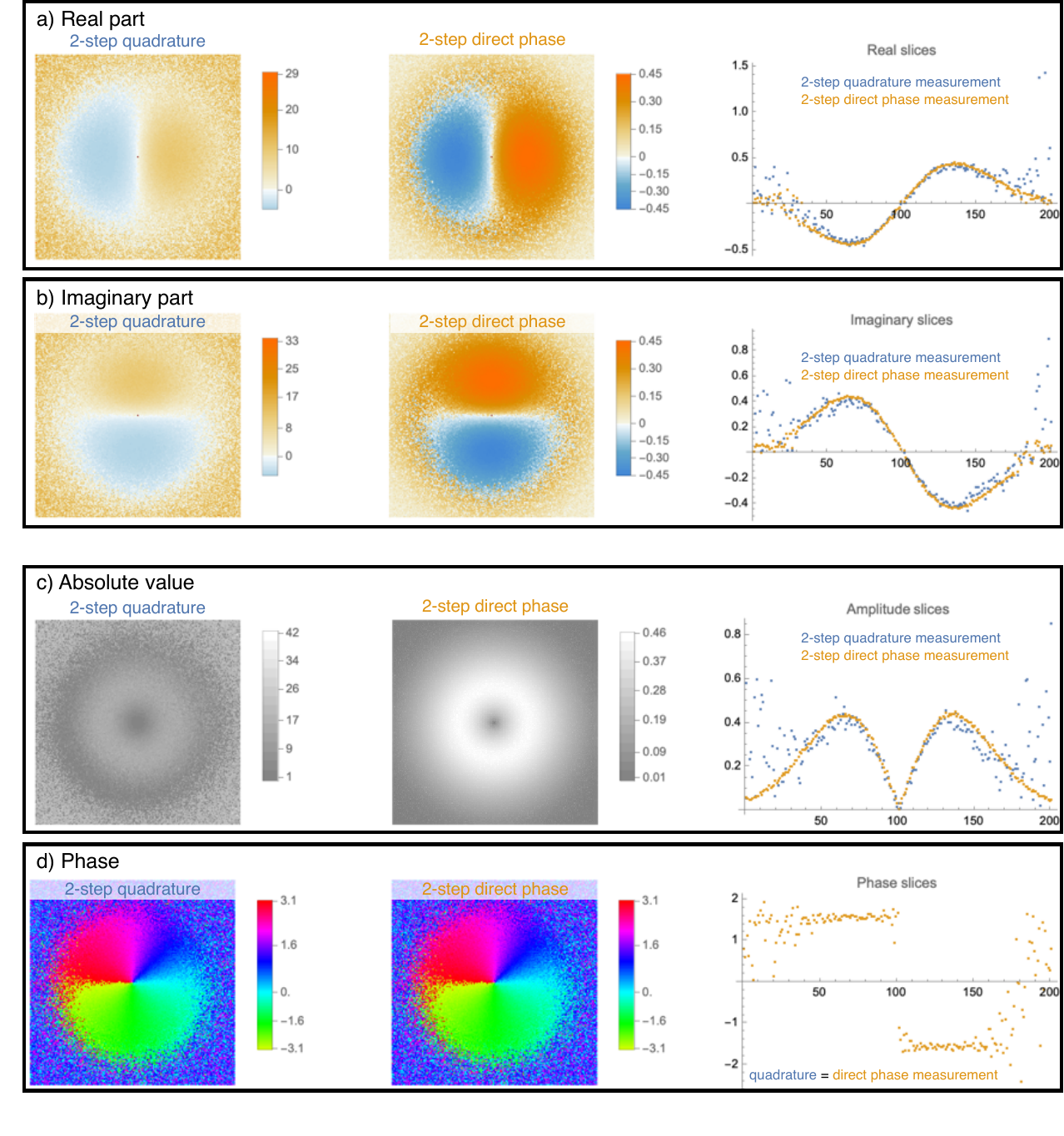}
    \caption{a) Real and b) imaginary, and c) absolute value and d) phase of phase-resolved images from 2-step quadrature PSI (left) and 2-step direct phase PSI (middle). The ground truth is the same as in Figure \ref{fig:ArbitraryVsQuadraturePSI}. 
    }
    \label{fig:QuadratureVsDirecPhasePSI}
\end{figure}

Using the same computational data as described above, we also implemented direct-phase PSI, following Equations 4 and 6 in the main paper. Figure \ref{fig:QuadratureVsDirecPhasePSI} shows a comparison of the resulting fields obtained by direct-phase vs quadrature PSI, for both 2D images and 1D slices along the $x$ axis for $y=0$. The two methods yield the same results for the phase, but the direct-phase technique is much less noisy in real, imaginary, and modulus parts of the data.

\section{SLM calibration}
\begin{figure}[hbt!]
    \centering
    \includegraphics[width=.8\linewidth]{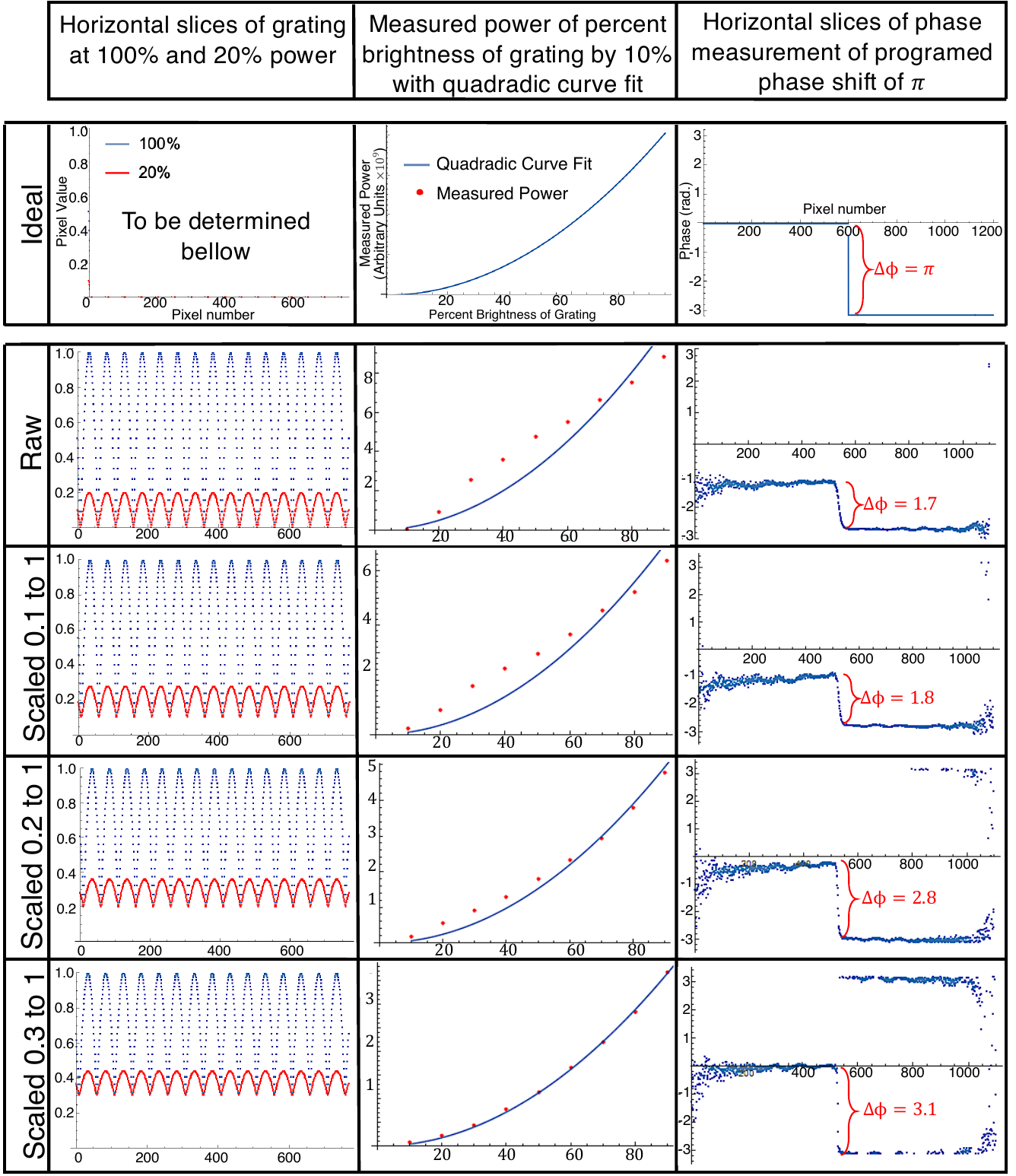}
    \caption{Horizontal slice of two gratings representing 100\% and 20\% brightness of a complex grating structure used to calibrate the SLM with corresponding characterization of power versus brightness with curve fit, and a horizontal slice of a phase measurement of a $\pi$ phase shift. }
    \label{fig:SLMCalibration}
\end{figure}

The SLM was calibrated by testing the linearity of its modulation on the light, then scaling and testing controlled gratings to correct for any non-linearity. As a starting point, a grating that scaled from 0 to 1 in grayscale amplitude was generated. Then this grating was multiplied by values ranging from 0.1 to 0.9 in increments of 0.1, to representing different levels of brightness within a more complex grating structure. Then the power of the resulting first-order diffraction was measured and plotted against the corresponding percent brightness.  Measurements are shown with arbitrary units of power that correspond to the sum of all pixel values of the measured image after background subtraction. When 100\% of the pixel range was used, the power was not scaled quadratically as expected. We expect the power to scale quadratically as brightness of the grating increases because the grating modulates the beam amplitude, and the measured intensity is proportional to the amplitude squared.

We tested the power when the top 90\%, 80\% and 70\% of the SLM's pixel brightness was used and again plotted each of these against percent brightness.  To verify the impact on the phase we implemented and measured a $\pi$ phase shift across half of the grating for each of the updated scales. The obtained results were consistent with the brightness graphs: when the power plot more closely fit the quadratic curve, the phase shift was closer to the expected $\pi$ phase shift. We found the phase shift, and quadratic curve, that most closely aligned with the expected value occurred when the gratings were scaled from 0.3 to 1. We then applied this to the experiment by scaling all gratings from 0.3 to 1: meaning the top 70\% of the SLM's pixels brightness was utilized. In the next section, we show results of implementing this for typical holographic gratings.

\section{Sample Gratings}

We used diffraction-grating holography to control the amplitude and phase of complex laser modes.To generate the holograms we first started by defining the Laguerre-Gaussian (LG) mode. To construct this we used the equation:

\begin{equation}
   \mathrm{LG} = \textrm{gaussian}(x, y) \times  \textrm{norm}(\ell, p) \\
   \times \textrm{azimuthal}(m, x, y) \times \textrm{core}(m, x, y) \times \textrm{radial}(\ell, p, x, y)
\end{equation}
Where:

\begin{equation}
    \textrm{gaussian}(x, y) = \exp\left[-\frac{x^2 + y^2}{w_0^2}\right] 
\end{equation}

\begin{equation}
    \textrm{azimuthal}(\ell, x, y) = \exp\left[i \, \ell \arctan\left(\frac{y}{x}\right)\right] 
\end{equation}

\begin{equation}
     \textrm{core}(\ell, x, y) = \left(\frac{\sqrt{2}}{w_0} \sqrt{x^2 + y^2}\right)^{|\ell|} = \left(\frac{2 (x^2 + y^2)}{w_0^2}\right)^{|\ell|/2} 
\end{equation}

 \begin{equation}
    \textrm{radial}(\ell, p, x, y) = \textrm{LaguerreL}(\ell, |m|, \frac{2 (x^2 + y^2)}{w_0^2}) 
\end{equation}
\begin{equation}
    \textrm{norm}(\ell, p) = \sqrt{\frac{2 p!}{\pi (p + |\ell|)!}}
\end{equation}

For the case $\ell=4$, $p=0$, and after a bit of simplification, the equation for the LG mode is:
\begin{equation}
     \mathrm{LG}(\ell=4,p=0,x,y)= \frac{1}{2\sqrt{3\pi}} \exp\left(-\frac{x^2 + y^2}{w_0^2}\right) \exp\left(4i \, \arctan\left(\frac{y}{x}\right)\right) \left(\frac{2 (x^2 + y^2)}{w_0^2}\right)^2
\end{equation}

Then the amplitude of this mode is generated using:
\begin{equation}
    \sqrt{I_S}=|\mathrm{LG}(\ell=4,p=0,x,y)|
\end{equation}

The tilted planewave is also generated using the equation:
\begin{equation}
    \exp[i \, ( \frac{\sqrt{3}}{2} x + \frac{1}{2} y )]
\end{equation}

The signal phase is generated using the equation: 
\begin{equation}
    \exp[i \, \textrm{arg}[\mathrm{LG}(\ell=4,p=0,x,y)]]
\end{equation}

These terms are then combined into an equation for the signal grating: 
\begin{equation}
    T_{S, \ell=4}=\sqrt{I_S}\times[\exp[i \, ( \frac{\sqrt{3}}{2} x + \frac{1}{2} y )]+\exp[i \, \, \textrm{arg}[LG(\ell=4,p=0,x,y)]]]
\end{equation}

The reference grating is generated using the equation:
\begin{equation}
    T_{R, \ell=4}(\delta)=[\exp[i \, \delta]+\exp[i \,( \frac{\sqrt{3}}{2} x + \frac{1}{2} y)]] \times \sqrt{\frac{2}{\pi}}\exp[\frac{x^2+y^2}{w^2}]
\end{equation}

The interference modes are then generated: 
\begin{equation}
  T_{R+S,\delta=0}= T_{S, \ell=4}+ T_{R}(\delta=0)
\end{equation}
\begin{equation}
  T_{R+S,\delta=\pi/2}= T_{S, \ell=4}+ T_{R}(\delta=\pi/2)
\end{equation}

Shown below in Figure S.4 is an example set of gratings for generating an LG $\ell=+4$ signal mode, along with the reference mode and interference modes with $\delta=0$ and $\delta=\pi/2$ that are needed for direct-phase PSI. S.4a shows gratings before calibration and S.4b shows the gratings that were calibrated using the procedure identified in section S.3. This procedure was used to make all of the gratings that were used to generate the modes that were measured in the paper.

\begin{figure}[hbt!]
\centering
    \includegraphics[width=.75\linewidth]{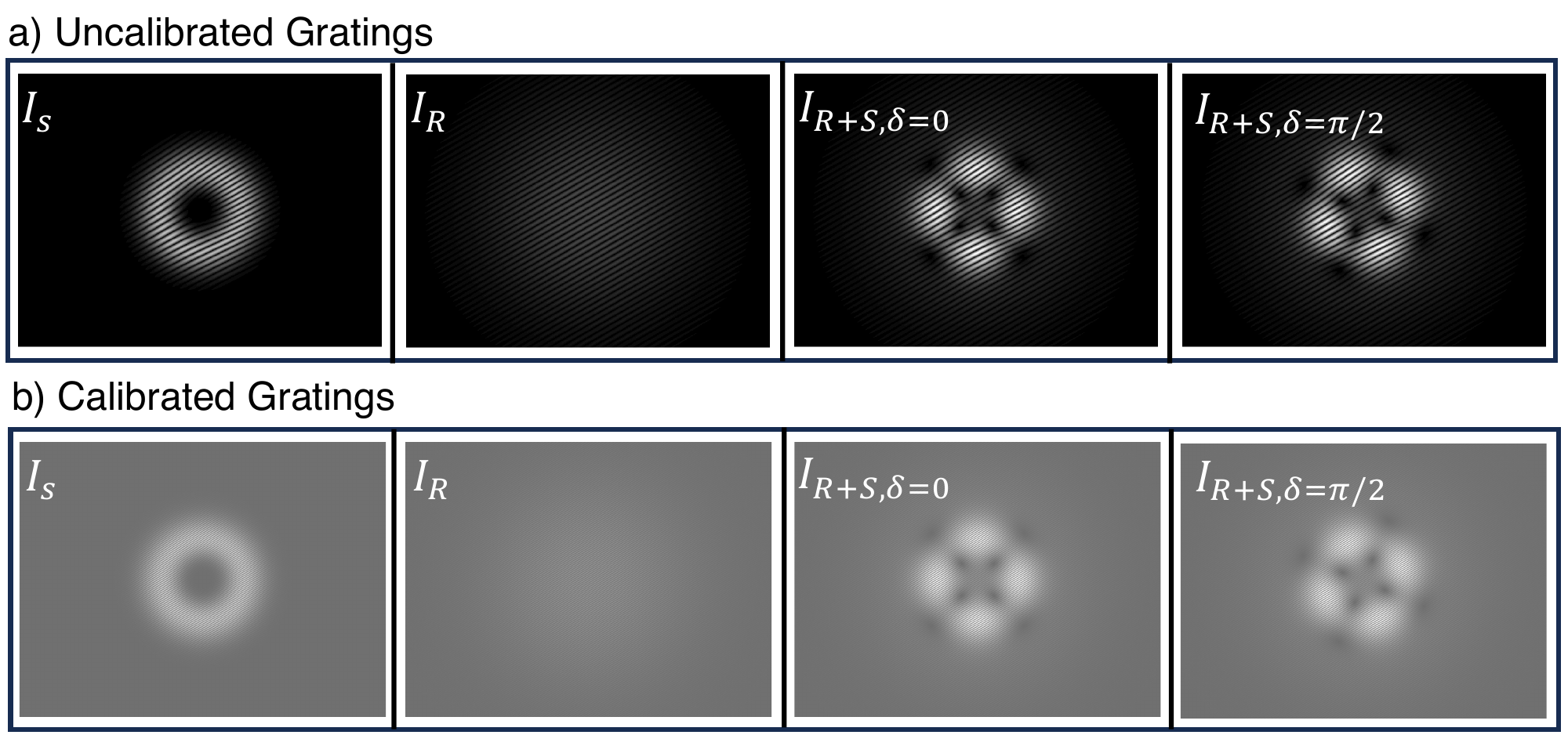}
    \caption{Sample gratings that produce an
$\ell$=4 Laguerre-Gaussian (LG) mode as shown in Figure 1. These include: the grating for the signal beam $I_S$, the reference beam $I_R$, the combined gratings $I_{R+S, \delta =0}$, and the shifted and combined gratings $I_{R+S, \delta =0}$. a) Depicts uncalibrated gratings and b) depicts gratings that are scaled to correct for the non-linearity of the SLM}
    \label{fig:Gratings}
\end{figure}

\section{Correcting Rotated Images}

In the phase-resolved images measured in Figure 4 
of the paper, we noted a slight tilt in the measured phases and amplitudes. We identified the tilt as coming from the beam not being a consistent height after the SLM. This issue is easily corrected, as shown in Figure S.5. Shown are two images recorded  when the SLM is programmed with a grating with a patchwork phase mask. The image on the left  was acquired with the original setup, in which the rotation of the image is attributed to a partial periscope effect from a changing beam height. The image on the right was taken following the correction of the height of the beam after the SLM. This demonstrates successful correction of the slight tilt. This partial periscope effect also explains the slight phase offset displayed in the line-outs of the phase image in Figure 1.

\begin{figure}
    \centering
    \includegraphics[width=.6\linewidth]{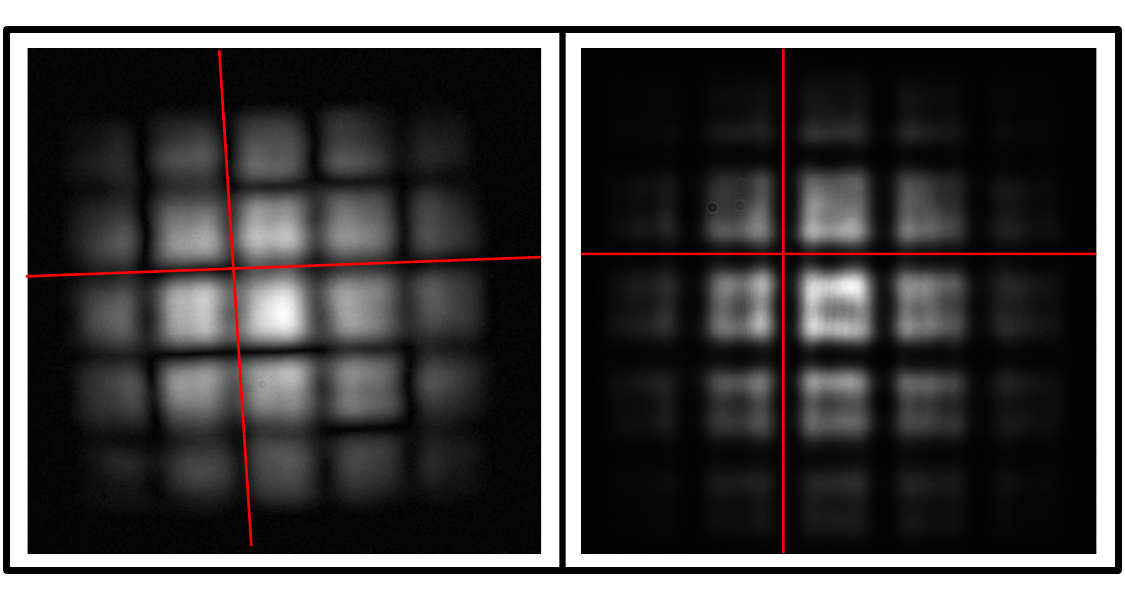}
    \caption{Comparison of recorded images before and after correcting beam height. Left: Original setup shows image rotation due to partial periscope effect. Right: Image after height adjustment, demonstrating successful tilt correction.}
    \label{fig:Updatedtilt}
\end{figure}

\end{document}